# A Scalable Trust Discovery Architecture for the Internet of Agents

**Song Zhang** [a, c], **Jiankang Yao** [a, c, *], **Hongtao Li** [a, c],
**Xiaojun Zhang** [b], **Xugang Shen** [a, c], **Xin Li** [a, c], **Yanbiao Li** [d]

[a] China Internet Network Information Center, Beijing 100070, China

[b] Alibaba Cloud Intelligence Group, Beijing 100102, China

[c] National Engineering Laboratory of Internet Domain Name Management Technology, Beijing 100070, China

[d] Computer Network Information Center, Chinese Academy of Sciences, Beijing 100083, China

* Corresponding author at: Building 4, No.9 West Road, Automobile Museum, Fengtai District, Beijing 100070, China. E-mail address: yaojk@cnnic.cn (J. Yao)

**Abstract:** The Internet of Agents is expected to enable large numbers of autonomous agents to discover, verify, and collaborate with each other across heterogeneous platforms. However, current agent protocols mainly address tool invocation and inter-agent communication, leaving scalable agent registration, trustworthy identification, and capability-oriented discovery largely unresolved. To address this, this paper proposes a scalable trust discovery architecture for the Internet of Agents. The proposed architecture adopts a hierarchical and distributed design consisting of three layers: Agent Root for trusted registry governance, Agent Registry for agent registration and metadata publication, and Agent Resolver for distributed capability discovery and trust-aware resolution. The architecture further introduces a registry-suffix-anchored composite identity scheme, which binds an agent's native identifier to a trusted registry suffix to generate a globally discoverable identity. It also incorporates a dual-certificate and multi-level authentication mechanism to strengthen identity trust among agents. We implement a prototype and evaluate it through large-scale agent registration and resolution experiments. The prototype achieves an average registration latency of 58ms and an average discovery latency of 25ms, and it supports more than 19,000 registration requests per second and more than 29,000 agent discovery requests per second. These results demonstrate the feasibility of the proposed architecture, providing a practical approach toward scalable and identity-trusted agent ecosystems in the Internet of Agents.

**Keywords**: Internet of Agents, Trust discovery architecture, Agent identity management, Multi-level authentication

## 1. Introduction

In recent years, artificial intelligence (AI) has been undergoing a fundamental transition from passive content generation toward autonomous task execution. Large language models (LLMs) have significantly enhanced the reasoning, planning, and natural language understanding capabilities of intelligent systems, making it possible to construct autonomous agents that can perceive user intent, decompose complex goals, interact with external tools, maintain contextual memory, and complete multi-step tasks with limited human intervention [1, 2]. LLM-based agents are commonly built around several core components, including planning, memory, tool use, environmental feedback, and task execution, which together enable them to move from simple dialogue systems to goal-oriented intelligent entities [3]. Compared with traditional rule-based agents or single-purpose software bots, LLM-based agents exhibit stronger generalization ability, richer interaction patterns, and more flexible decision-making capabilities [4].

This evolution has also accelerated the emergence of agent-oriented software ecosystems. Instead of being isolated applications, agents are increasingly expected to connect with external services, data sources, tools, and other agents. Representative studies such as ReAct [3] and Toolformer [4] demonstrate that LLMs can improve task execution by combining reasoning with external actions or by learning when and how to invoke tools through APIs. Moreover, generative agents and embodied agents further show that LLM-based agents can maintain memory, simulate social behaviors, acquire skills, and perform long-horizon tasks in interactive environments [5, 6].

As agent ecosystems continue to expand, the concept of the Internet of Agents (IoA) is becoming an important vision for future intelligent infrastructure. Massive numbers of agents may be deployed by individuals, enterprises, communities and organizations to provide specialized services, such as financial analysis [7], medical consultation [8], software development [9], etc. These agents will not merely respond to user commands; they may actively search for suitable collaborators, invoke remote capabilities, negotiate task execution, and dynamically form multi-agent workflows [10]. Recent studies on AI agent protocols and IoA architectures have emphasized that future agent ecosystems require standardized mechanisms for identity management, capability description, discovery, interoperability, and cross-domain collaboration [11-13].

However, the large-scale deployment of intelligent agents also introduces a series of fundamental infrastructure challenges. First, agent identity management becomes difficult when agents are created, updated, migrated, or terminated across heterogeneous platforms and administrative domains. Without a unified naming and registration mechanism, it is hard for one agent to accurately identify another agent or verify whether the discovered agent is legitimate [12, 14]. Second, capability discovery becomes increasingly complex for agents may describe their functions in natural language, expose different service interfaces, or update their capabilities over time. Simple keyword matching or platform-specific directories are insufficient for supporting scalable and semantically accurate agent discovery in large-scale IoA environments [12, 13]. Third, trust and security become critical concerns. Malicious agents may publish false capability claims, impersonate legitimate agents, poison discovery indexes, or exploit inter-agent communication channels. Recent studies have shown that LLM-based agents pose new security and privacy risks due to their autonomy, tool-use ability, memory mechanisms, and interaction with external environments expand the attack surface beyond conventional LLM applications [15, 16].

Therefore, future agent infrastructure requires more than communication protocols between already-known agents. It needs a foundational discovery and governance system that can answer three basic questions: who the agent is, what the agent can do, and whether the agent's identity can be verified and trusted. To achieve these, we propose a scalable trust discovery architecture for IoA. Inspired by the hierarchical delegation and resolution principles of the Domain Name System (DNS), the proposed architecture introduces three key functional entities: Agent Root, Agent Registry, and Agent Resolver, as shown in Figure 1. Specifically, Agent Root serves as the global trust anchor for managing authorized registries, Agent Registry supports agent registration and capability card management, and Agent Resolver enables distributed agent discovery, capability retrieval, and trust-aware resolution. Through this design, the proposed architecture provides a scalable approach for agent identity management, capability discovery, and trustworthy inter-agent connection. The main contributions of this paper are as follows:

1. We propose a scalable trust-aware agent discovery architecture, that adopts a Root–Registry–Resolver architecture to support trusted registry management, agent registration, agent resolution and capability-oriented discovery, thereby addressing agent identity management and trustworthy identification in large-scale IoA environments.
2. We introduce a registry-suffix-anchored composite identity scheme that binds heterogeneous agent identifiers with trusted registry namespaces, enabling verifiable registration and discovery.

3. We design a dual-certificate and multi-level authentication mechanism to enhance agent identity trust. Private and public certificate authority (CA) mechanisms support efficient intra-domain and trustworthy cross-domain agent interconnection, while multi-level authentication enhance agent identity trust.
4. We implement a prototype of the proposed architecture and evaluate its performance. The prototype achieves 58ms registration latency, 25ms discovery latency, and supports over 19,000 registration requests per second and 29,000 agent discovery requests per second, demonstrating its effectiveness and scalability.

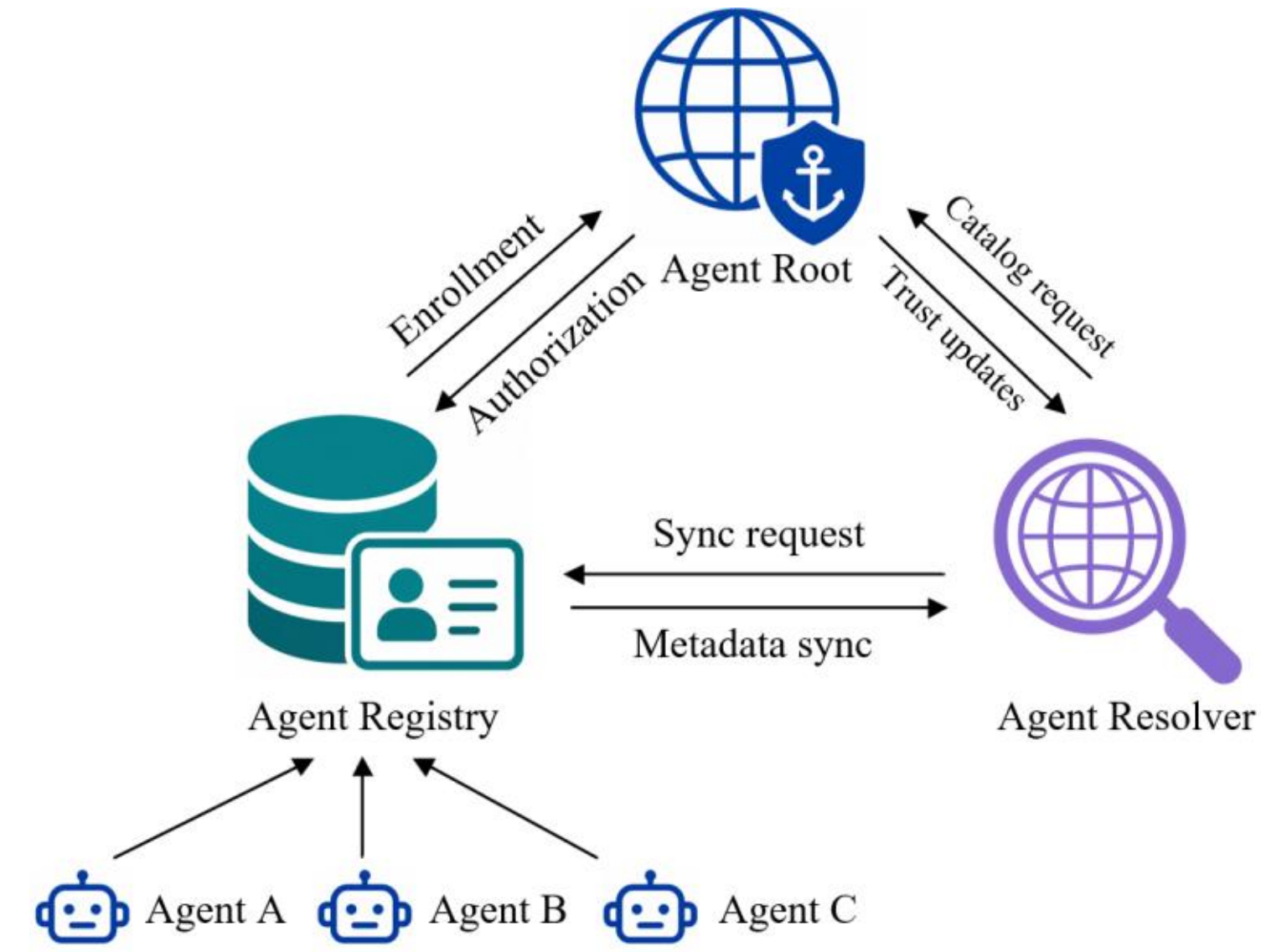

Figure 1. Simplified schematic of the proposed discovery architecture. The architecture adopts a three-tier Root–Registry–Resolver architecture. The Root manages trusted registration, the Registry handles agent registration and information publication, and the Resolver synchronizes trusted data and processes capability queries.

## 2. Related works

LLM-based agents have extended the autonomy and task-execution capabilities of AI systems, thereby accelerating the development of infrastructure tailored to intelligent agents. This section reviews related work from two perspectives: the development of intelligent agents, the DNS and its application to agent systems.

### 2.1 LLM-based agents

LLM-based agents have evolved from simple dialogue interfaces into autonomous systems capable of reasoning, planning, tool use, memory management, and environment interaction. He et al. [17] proposed an end-to-end multimodal web agent that interacts with real-world websites through visual perception and action execution, demonstrating the ability of LLM-based agents to operate in open web environments. Xi et al. [18] introduced AgentGym, a framework that supports agent exploration, training, and evaluation across diverse interactive environments, further showing that agents can evolve through environmental feedback and self-improvement. Meanwhile, recent multi-agent frameworks such as AgentScope [19], Magentic-One [20], and MARCO [21] indicate that complex tasks can be decomposed and coordinated through message exchange, specialized agent roles, orchestration agents, and workflow-level control. These studies suggest that the development of intelligent agents is shifting from improving the capability of individual models toward constructing organized agent ecosystems, where heterogeneous agents can cooperate, specialize, and participate in larger service workflows.

As the scope of agent capabilities continues to expand, agent-oriented protocols have emerged to support tool integration and cross-agent collaboration. Model Context Protocol (MCP) standardizes how LLM applications connect to external data sources, tools, and workflows, thereby addressing the problem of agent-to-tool interaction [22]. In contrast, the Agent2Agent (A2A) protocol focuses on agent-to-agent communication and aims to enable heterogeneous agents to exchange information, delegate tasks, and coordinate execution across different platforms [23]. Recent surveys on AI agent protocols have further compared MCP, A2A, Agent Network Protocol (ANP) [24], Agora [25], and other protocol designs, showing that protocol-level interoperability is becoming a key enabler of future multi-agent ecosystems [11].

### 2.2 DNS for agents

DNS is one of the most successful distributed infrastructures on the Internet. Traditional DNS defines a hierarchical namespace and a distributed resolution mechanism that maps human-readable domain names to network resources [26]. Its tree-structured naming model, delegation mechanism, caching strategy, and distributed authority make it highly scalable and robust for global Internet deployment. Beyond basic name-to-address resolution, DNS has also been extended to support service discovery. DNS-Based Service Discovery (DNS-SD) defines how DNS resource records can be used to discover named service instances within a domain, while related DNS-based registration mechanisms further support dynamic service publication and discovery [27]. These mechanisms indicate that DNS is not only a naming system, but also a general-purpose infrastructure for distributed service discovery.

Recently, DNS-inspired designs have begun to be explored for intelligent agent discovery. Cui et al. [28] proposed a root-domain naming and service discovery system for LLM agents, AgnentDNS, aiming to support cross-vendor agent and tool service registration, semantic service discovery, secure invocation, and unified billing. Huang et al. [14] introduced a DNS-based universal directory for secure AI agent discovery and interoperability. Akshay and Elyson [29] proposed a conceptual trust layer built based on DNS to support agent discovery, identity management, and governance. These works demonstrate the feasibility of applying DNS principles to agent naming, discovery, and trust management. However, further improvements are still needed in trusted registry management, standardized capability representation, resolver-side capability indexing, and governance-oriented revocation. To address these gaps, this paper proposes a scalable trust discovery architecture by integrating DNS-inspired hierarchical discovery with agent capability publication and root-governed trust management.

## 3. Proposed method

### 3.1 Overview of the proposed architecture

Inspired by the hierarchical delegation and resolution principles of the DNS, we propose a scalable trust discovery architecture for the IoA. The proposed architecture mainly consists of three functional layers: the top trust layer (Agent Root), the middle distributed layer (Agent Registry), and the bottom discovery layer (Agent Resolver), as shown in Figure 2. Agent Root acts as the global identity trust anchor. It maintains the trusted registry suffix catalog, validates the enrollment of Agent Registries, distributes trusted registry information to Agent Resolvers, and broadcasts revocation information when a registry becomes compromised or malicious. Agent Registry is responsible for agent registration and metadata publication. An Agent Registry owns a fixed registry suffix and accepts registration, update, and unregistration requests from agents or domain owners. Agent Resolver provides the discovery entry point for agent clients. It periodically obtains the trusted registry catalog from Agent Root and synchronizes agent metadata from valid Agent Registries.

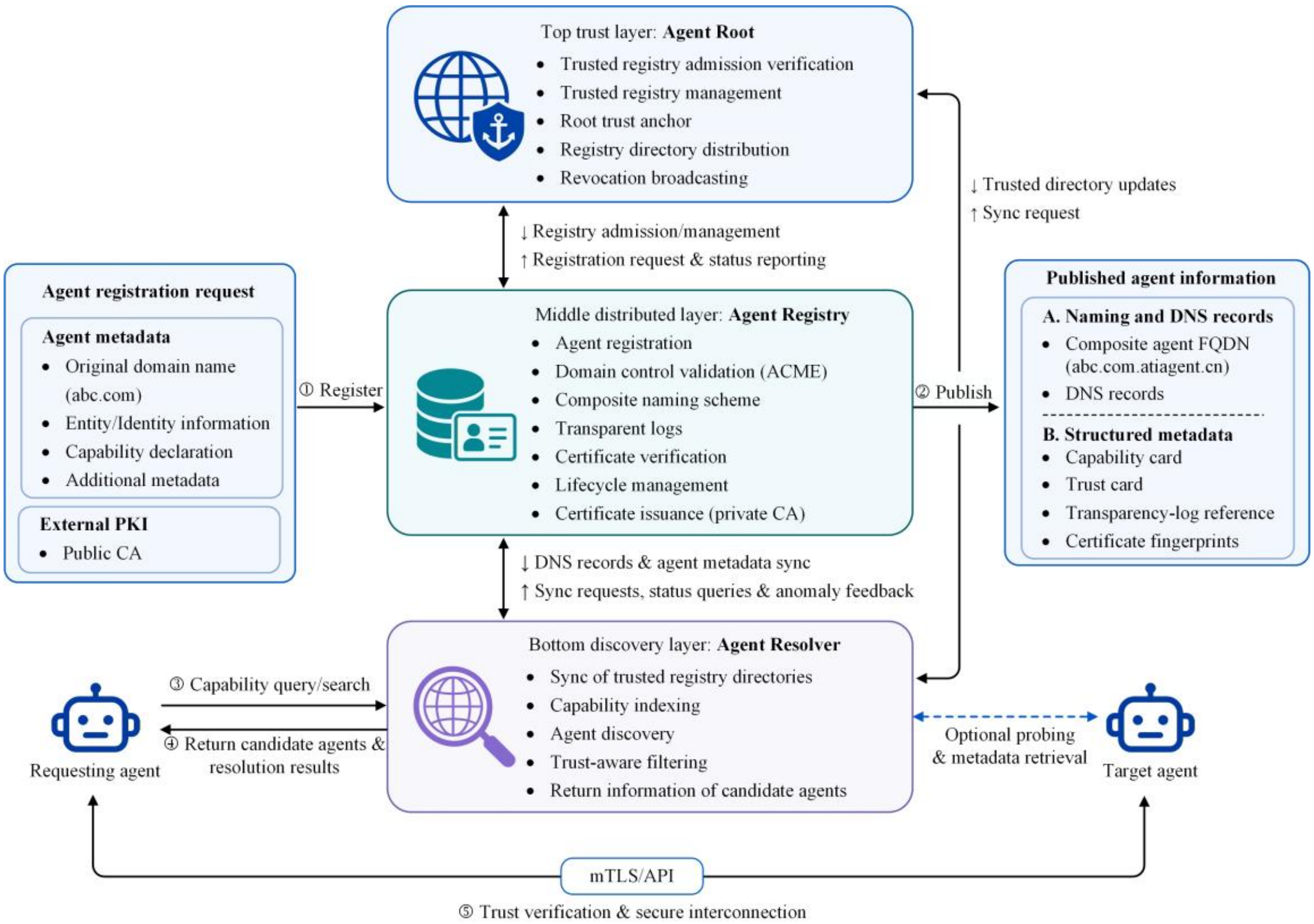

Figure 2. The architecture diagram of the proposed architecture.

The workflow of the proposed architecture consists of five main stages. First, an agent or domain owner registers with an Agent Registry, where domain-control validation and identity binding are performed. Second, the Agent Registry publishes DNS records and structured metadata, including agent identity, capability, and trust information. Third, when a requesting agent needs to find a target capability, it submits a capability-oriented query to the Agent Resolver. Fourth, the Agent Resolver searches its local capability index, filters invalid or untrusted records, and returns a list of candidate agents with their corresponding metadata. Finally, the requester verifies the returned trust evidence and establishes a secure connection with the selected target agent through mechanisms such as mTLS, DANE, DNSSEC, and transparency log validation.

Through this design, the proposed architecture separates global trust management, registration management, and discovery resolution into distinct but coordinated layers. Compared with a centralized directory, the Root–Registry–Resolver model improves scalability by distributing agent metadata across registries and resolvers. Compared with ordinary communication protocols that assume a known peer, the proposed architecture provides an infrastructure-level discovery process that allows agents to be identified, discovered by capability, verified, and connected in a trustworthy manner.

### 3.2 Composite agent identity

A key challenge for Internet-scale agent discovery is not the absence of agent identifiers, but the lack of a globally verifiable trust anchor that binds heterogeneous agent identifiers to authorized registration domains. To address this challenge, we introduce a registry-suffix-anchored composite identity scheme. The key idea is to preserve the agent's native identifier as its registration prefix while binding it to a trusted registry suffix managed by authorized Agent Registry. The native identifier is not limited to a conventional domain name. It can be an original domain, a platform-assigned agent ID, a decentralized identifier (DID), a public-key fingerprint, or another verifiable identifier used by an agent platform or organization. As shown in Figure 3, the composite identity scheme normalizes the native identifier into a DNS-compatible registration prefix and then combines it with a trusted registry suffix. Formally, given an agent native identifier $I_a$ and a trusted registry suffix $S_r$, this scheme generates a composite agent identity as $F_a = Normalize(I_a).S_r$, where $F_a$ denotes the globally discoverable composite fully qualified domain name (FQDN) of the registered agent. For example, an agent associated with the native domain `abc.com` can be registered under the trusted registry suffix `atiagent.cn`, resulting in the composite identity `abc.com.atiagent.cn`.

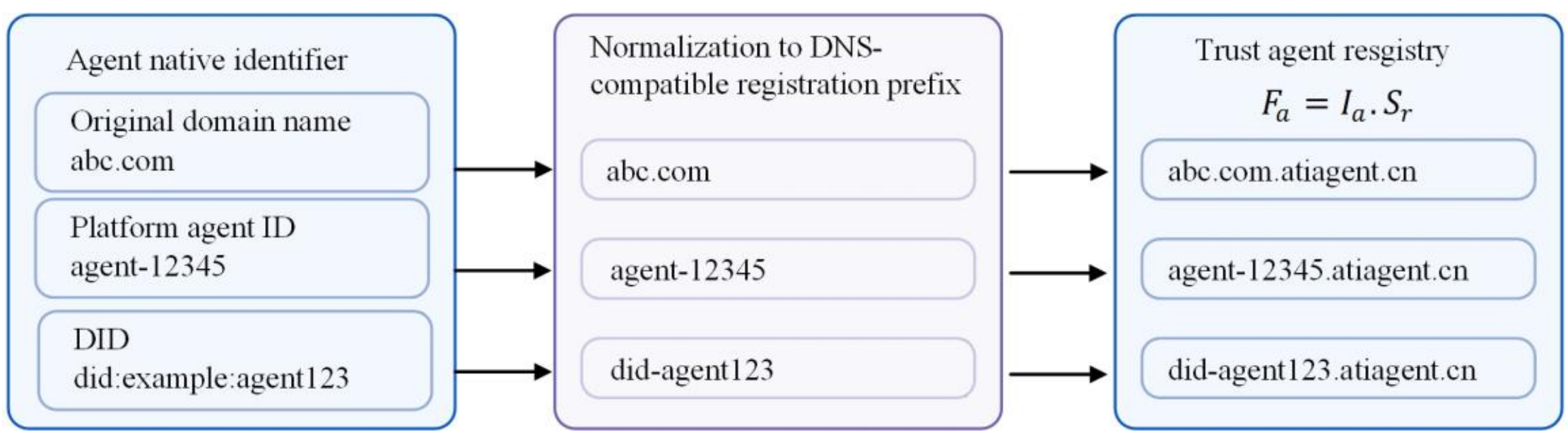

Figure 3. Registry-suffix-anchored composite identity establishment.

This scheme transforms independently operated agent identifiers into globally verifiable registered identities by attaching a trusted registry namespace. The composite identity does not replace the native identifier; instead, it extends the identifier with a verifiable registry context. During registration, the registry verifies domain control, generates the composite FQDN under its fixed suffix, publishes DNS records, and hosts the associated capability card or trust card. For domain-based identifiers, this verification can be performed through domain-control validation such as ACME. For non-domain identifiers, the registry can use platform-side verification, public-key challenge-response, DID-based verification, or organizational identity review. Since all agents under the same registry suffix share a unified namespace, resolvers can synchronize, index, and search agent metadata in a structured manner.

Compared with approaches that treat each agent domain as an isolated identifier, the registry-suffix-anchored design improves scalability, governance, and interoperability. Agent Root only needs to maintain the trusted registry suffix catalog, while Agent Registries manage their own namespaces and Agent Resolvers perform discovery based on composite identities and capability descriptions. Because the composite identity is compatible with the DNS hierarchy, it can be published, resolved, cached, and synchronized using mature DNS mechanisms. Therefore, this scheme provides the foundation for scalable agent registration, capability-oriented discovery, lifecycle management, and trust verification.

### 3.3 Dual-certificate and multi-level authentication

To ensure that a discovered agent is not only reachable but also identity-verifiable, the proposed architecture introduces a dual-certificate authentication mechanism. As shown in Figure 4. In this mechanism, the private CA is mainly used to support fast and controllable trust establishment among agents within the same trust domain. Since these agents are usually managed by the same organization or platform, the private CA can provide lightweight identity issuance, local lifecycle management, and efficient intra-domain identity verification. In contrast, the public CA is mainly used to enable reliable interconnection among agents across different trust domains. The public CA authorizes the intermediate CA, which then issues certificates to the agent within their corresponding trust domains. This design inherits the global trust model of PKI while delegating agent identity issuance, governance, and risk management to a controllable intermediate layer.

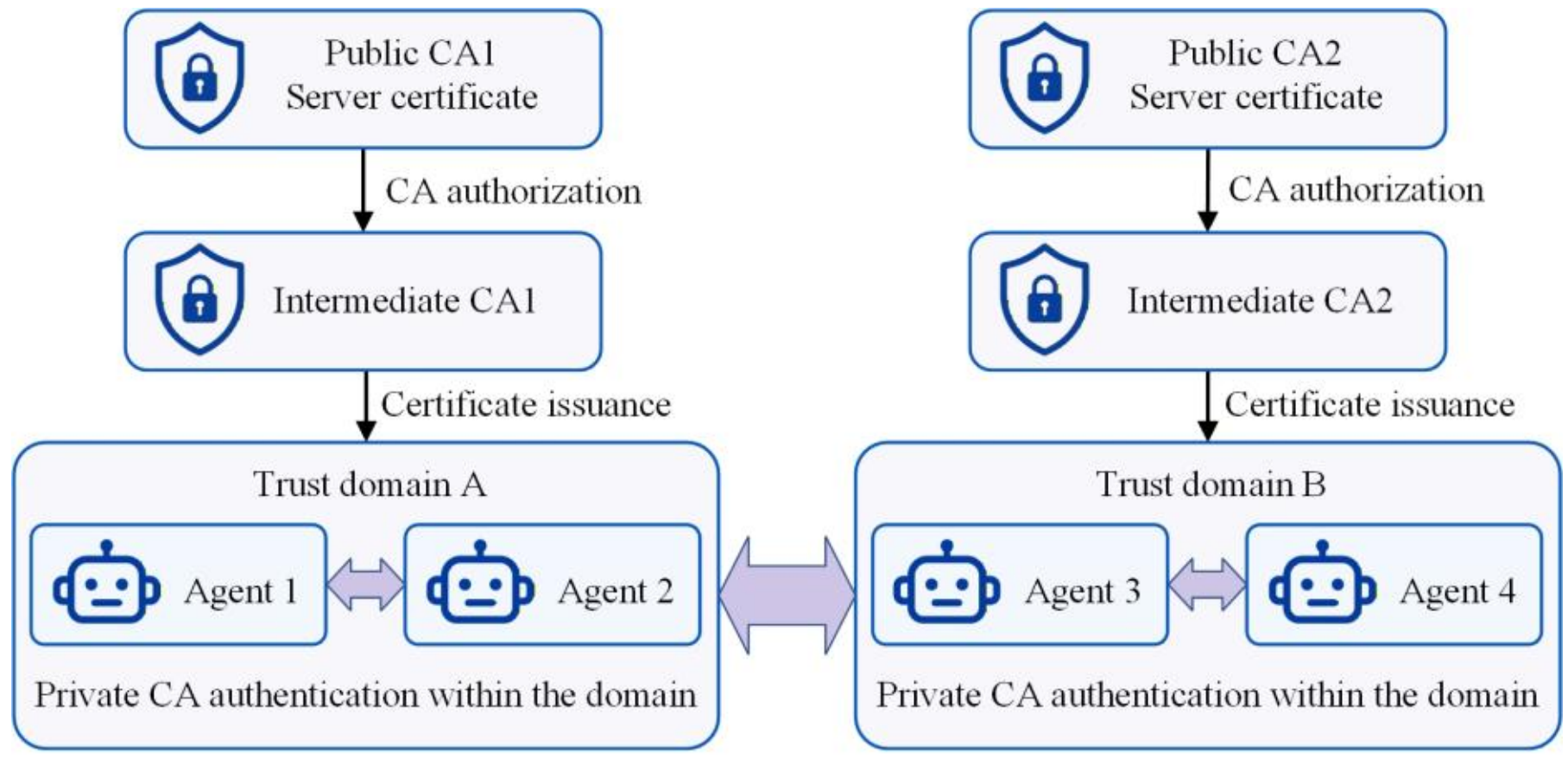

Figure 4. Dual CA authentication mechanism.

Based on this dual-certificate mechanism, the proposed architecture defines three progressive authentication levels: Basic Authentication, Enhanced Authentication, and Advanced Authentication. These levels correspond to different trust requirements and allow agent clients to select an appropriate verification strength according to the application scenario. As shown in Table 1, Basic Authentication verifies both the public CA server certificate and the private CA identity certificate, and establishes an mTLS-secured channel to confirm the peer agent's communication security and registered identity. Enhanced Authentication further checks transparency log records to ensure that the agent identity and certificate information are traceable and have not been silently modified. Advanced Authentication additionally uses DANE and DNSSEC to verify the binding between the agent domain, DNS records, and certificate public key. This progressive design enables flexible trust enforcement for scenarios ranging from lightweight service discovery to high-assurance inter-agent collaboration.

Table 1. Multi-level authentication mechanism.

| Authentication | Main mechanisms | Main function |
| --- | --- | --- |
| Basic authentication | Public CA certificate, private CA identity certificate, mTLS | Communication channel and registered agent identity, confirms that the peer agent owns valid communication and identity credentials. |
| Enhanced authentication | Transparency log verification, certificate fingerprint checking | Registration event, certificate fingerprint, lifecycle record, makes agent registration auditable, traceable, and tamper-evident. |
| Advanced authentication | DANE, DNSSEC, certificate public-key binding | Binding among domain name, DNS record, and certificate public key，strengthens the end-to-end trust chain from DNS identity to certificate identity. |

### 3.4 Capability publication and trust-aware

In addition to identity registration and authentication, the architecture must enable agents to be discovered according to their functional capabilities. To this end, Agent Registry publishes structured metadata for each registered agent through capability cards or trust cards. A capability card describes the agent's composite identity, DNS service records, network endpoints, certificate information, online status, registration expiration time, and capability descriptions. A trust card further extends this metadata with identity certificate fingerprints, transparency log references, lifecycle status, and trust level information. By standardizing agent capability and trust metadata, the architecture allows heterogeneous agents to be described, published, synchronized, and interpreted in a consistent manner.

Agent Resolver performs trust-aware capability discovery based on the metadata synchronized from trusted Agent Registries. It first obtains the trusted registry catalog from Agent Root, then synchronizes capability cards or trust cards from valid registries and builds local indexes over composite identities, capability descriptions, and trust attributes. When a requesting agent submits a capability query, the resolver searches the local capability index, filters expired, revoked, or untrusted agents, and returns candidate agents together with their corresponding metadata. The requester can then perform multi-level authentication before establishing a secure connection. In this way, the proposed architecture transforms agent discovery from simple name resolution into a capability-oriented and trust-aware resolution process.

## 4 Prototype evaluation and case study

### 4.1 Setup

To evaluate the feasibility of the proposed architecture, we implemented a prototype system named ATI (agent trust infrastructure) following the three-layer architecture. The prototype consists of five core components: a Nginx reverse proxy, an Agent Root service, Agent Registry services, transparency log services, and Agent Resolver services.

The Nginx reverse proxy is deployed as the unified access entry point for external requests and forwards communication traffic among different ATI components. The Agent Root service is responsible for maintaining the trusted registry information and providing global identity trust management functions. The Agent Registry service manages agent registration information, namespace allocation, and related metadata. The transparency log service records registration events, trust-related information, and lifecycle changes to provide auditable and traceable records. The Agent Resolver service synchronizes agent metadata and performs agent discovery and resolution operations based on the information maintained by the upper-layer components.

Each component is deployed on an independent server with same configurations. Each server is equipped with two Intel E5-2650 CPUs and 32 GB DDR3 memory.

### 4.2 Case study

We conducted an agent registration and discovery testing on Alibaba Cloud's Tongyi Qianwen agent platform, showing how an agent obtains a globally discoverable identity, publishes its capability information, and establishes trusted communication with other agents through the ATI.

During the agent registration phase, the agent client first generates a local key pair $(PK_a, SK_a)$ and submits a registration request containing the native agent identifier $I_a$, registration metadata $M_{reg}$, and the public key $PK_a$. The registration metadata includes the agent's name, version, service endpoint, supported interaction protocol, and capability descriptions. After receiving the request, the Agent Registry performs ACME DNS-01 domain-control validation to confirm that the requester has control over the associated domain. If the validation succeeds, the registry verifies and normalizes the self-declared registration metadata into published metadata $M_{pub}$, generates the composite agent identity $F_a = Normalize(I_a).S_r$, where $S_r$ is a Root-authorized trusted registry suffix. In this design, the Agent Root does not maintain per-agent registration data; instead, it maintains trusted registry namespaces and enables resolvers to identify which registries can be trusted. The registry then binds the identity, public key, version information, and dual-certificate credentials. The registry then generates the trust materials required for the supported authentication levels, appends key registration evidence to the transparency log, and publishes the corresponding DNS records, capability card, and trust card. As a result, the registered agent obtains a globally unique and DNS-compatible identity, while external entities can independently verify its ownership, version, public key, and life-cycle status through DNS records, certificate information, and transparency log evidence. The registration procedure is shown in Table 2.

Table 2. The procedure of agent registration

| Procedure 1: | Agent Registration and Trust Establishment |
|---|---|
| Input: | Agent native identifier $I_a$, Registration metadata $M_{reg}$ |
| Output: | Registration result |
| 1 | Agent client generates a key pair $(PK_a, SK_a)$ |
| 2 | Agent client submits registration request containing $I_a$, $M_{reg}$, $PK_a$ |
| 3 | Agent Registry performs ACME DNS-01 domain-control validation:<br>if validation fails:<br>Reject registration request<br>return |
| 4 | Agent Registry verifies and normalizes the registration metadata,<br>generate published metadata: $M_{pub} = Normalize(M_{reg})$ |
| 5 | Generate composite identity under the Root-authorized suffix:<br>$F_a = Normalize(I_a).S_r$, where $S_r$ is the trusted registry suffix. |
| 6 | Agent Registry binds $F_a$, $PK_a$, version, and dual-certificate credentials |
| 7 | Agent Registry generates the trust materials required for the supported authentication levels. |
| 8 | Agent Registry appends key registration evidence to the transparency log and publishes DNS records, capability card, and trust card. |
| 9 | Return registration result: $F_a$, $M_{pub}$, certificates, capability card, trust card, DNS records, log reference |

In the agent discovery phase, the Agent Resolver synchronizes the trusted registry catalog from the Agent Root. Based on this catalog, it synchronizes agent metadata from trusted Agent Registries and updates its local capability and identity indexes. Then the requesting agent submits a capability query or a target composite identity to the Agent Resolver. The resolver searches its local capability and identity indexes and retrieves candidate agents together with their associated metadata, including service endpoints, supported interaction protocols, capability cards, trust cards, certificate information, DNS records, and transparency log references. Before returning or using the discovery result, ATI filters candidate agents according to registry trust status, lifecycle state, expiration time, and the required authentication level. The requesting agent then verifies the returned trust evidence, including certificate credentials, domain or identity bindings, and transparency log records. Once verification succeeds, the requester can establish a secure communication channel with the selected target agent, for example through mTLS. The discovery procedure is shown in Table 3.

Table 3. The procedure of agent discovery and resolution

| Procedure 2: | Agent Discovery and Secure Interaction |
|---|---|
| Input: | Capability query $Q_c$ or target composite identity $F_a$ |
| Output: | Verified target agent and secure communication channel |
| 1 | Agent Resolver synchronizes the trusted registry catalog from Agent Root. |
| 2 | Agent Resolver synchronizes agent metadata based on trusted Agent Registries and updates its local capability and identity indexes. |
| 3 | Requesting agent submits a discovery query to Agent Resolver |
| 4 | Agent Resolver searches local capability or identity indexes and retrieves candidate agents. |
| 5 | Agent Resolver filters invalid candidates according to lifecycle status, expiration time, registry trust status, and supported authentication level. |
| 6 | Agent Resolver returns candidate agent results to the requesting agent. |
| 7 | Requesting agent verifies the returned trust evidence according to the required authentication level. |
| 8 | If verification fails:<br>    Reject the candidate agent<br>    return |
| 9 | Requesting agent establishes an mTLS-based secure connection with the verified target agent. |

### 4.3 Registration and resolution performance

To evaluate the efficiency of the proposed ATI prototype, we conducted performance tests for key operations using the wrk[1] benchmarking tool. The tested operations include agent registration, agent query, agent data publication, and agent review. Agent registration measures the end-to-end latency for submitting registration information, performing validation, generating the registration result, and preparing trust-related materials. Agent query measures the latency for resolving a target agent or a capability query through the Agent Resolver. Agent data publication measures the latency for publishing agent-related DNS records, whereas agent review measures only the time consumed by naming review.

[1] https://github.com/wg/wrk

Table 4 shows the latency measurements of ATI operations. Most operations can be completed within tens of milliseconds. Registration introduces higher latency than query and publication operations because it involves multiple trust-establishment steps, including domain-control validation, metadata processing, certificate binding, and transparency log recording. In contrast, agent query mainly relies on resolver-side local indexes, which reduces the cost of capability-based lookup and identity resolution. The relatively higher 99th %lie latency of agent query indicates the existence of tail latency under concurrent workloads, but the average and 95th %lie results still show that the resolver can provide low-latency discovery in most cases.

Table 4. The procedure of agent discovery and resolution

| Operation | Mean | 95th %lie | 99th %lie |
|---|---|---|---|
| Agent registration | 58ms | 127ms | 153ms |
| Agent discovery | 25ms | 32ms | 165ms |
| Agent data publication | 9ms | 10ms | 11ms |
| Agent review | 4ms | 6ms | 7ms |

We further measured the sustainable throughput of the prototype under concurrent requests. Without observable degradation in response latency, the prototype supports more than 19,000 QPS for agent registration and more than 29,000 QPS for agent query. These results indicate that ATI can support efficient agent registration and high-throughput agent discovery in the tested deployment.

### 4.4 Security analysis and considerations

The proposed architecture separates trust authorization, registration and publication, and discovery. This design establishes a verifiable trust chain, where Agent Root authorizes trusted registries, Agent Registries register and publish agent information, and Agent Resolvers perform trust-aware discovery. In addition, the architecture adopts a dual-certificate verification mechanism based on public and private CAs to bind an agent's network endpoint, composite identity, public key, and version information. Built on this dual-certificate mechanism, the architecture further defines three progressive levels of identity authentication, providing a flexible and reliable approach to trust establishment.

The proposed architecture mainly addresses identity-level trust and infrastructure-level governance, but it does not by itself guarantee the semantic correctness, service quality, or benign behavior of an agent after registration. A valid identity within the architecture indicates that the agent's ownership, certificate credentials, published metadata, and lifecycle status are verifiable, but a correctly registered agent may still produce unsafe outputs, misuse tools, or violate application-specific policies. Therefore, practical deployments should combine the proposed architecture with higher-layer mechanisms such as behavior monitoring, policy enforcement, risk scoring, and content or tool-use auditing, so as to enable safer and more controllable agent usage.

## 5 Conclusion

This paper proposes scalable trust discovery architecture for identity-trusted agent discovery and identity management in the Internet of Agents. The proposed architecture adopts a hierarchical Root–Registry–Resolver architecture to separate global trust governance, agent registration, and distributed discovery. Based on this architecture, we introduce a registry-suffix-anchored composite identity scheme for globally discoverable agent identities, and a dual-certificate multi-level authentication mechanism for verifiable and auditable agent identity. Through these designs, the proposed architecture enables trustworthy agent registration, capability-oriented discovery, identity verification, and inter-agent interaction. A prototype system named ATI was implemented to validate the feasibility and scalability of the proposed architecture. Experimental results demonstrate that the architecture can support large-scale agent registration and discovery with low latency and high throughput.

Despite its potential, the proposed architecture is still at an early stage of system design and prototype validation. Future work will focus on large-scale deployment experiments, privacy-preserving discovery, automated trust evaluation, and more comprehensive governance mechanisms for agent lifecycle management.

## CRediT authorship contribution statement

**Song Zhang:** Writing-original draft, Visualization, Validation, Methodology, Conceptualization. **Jiankang Yao:** Writing–review & editing, Supervision, Methodology, Funding acquisition, Conceptualization. **Hongtao Li:** Writing–review & editing, Funding acquisition. **Xiaojun Zhang:** Software, Writing–review & editing. **Xugang Shen:** Software, Validation. **Xin Li:** Software, Validation. **Yanbiao Li:** Writing–review & editing.

## Declaration of competing interest

The authors declare that they have no known competing financial interests or personal relationships that could have appeared to influence the work reported in this paper.